\documentclass{Interspeech}
\usepackage{multirow}
\usepackage{amsmath,graphicx}
\usepackage{xcolor}
\usepackage{blindtext}
\usepackage{multirow}
\usepackage{cleveref}
\usepackage{lipsum}
\usepackage{amsthm}
\usepackage{enumitem}
\usepackage{etoolbox}
\usepackage{graphicx}
\usepackage{makecell}
\usepackage{caption}
\usepackage{wrapfig}
\usepackage{setspace}
\usepackage{subcaption}
\usepackage{floatflt}
\usepackage{float}
\usepackage{setspace}
\usepackage{tikz}

\interspeechcameraready

\renewrobustcmd{\bfseries}{\fontseries{b}\selectfont}
\renewrobustcmd{\boldmath}{}
\newrobustcmd{\B}{\bfseries}
\newcommand\blfootnote[1]{%
	\begingroup
	\renewcommand\thefootnote{}\footnote{#1}%
	\addtocounter{footnote}{-1}%
	\endgroup
}
\newsavebox\CBox

\makeatletter

\renewcommand{\section}{\@startsection
   {section}%
   {1}%
   {}%
   {-0.5\baselineskip}%
   {0.2\baselineskip}%
   {}}%

\renewcommand{\subsection}{\@startsection
  {subsection}%
  {2}%
  {}%
  {-0.1\baselineskip}%
  {0.1\baselineskip}%
  {}}%

\renewcommand{\subsubsection}{\@startsection
  {subsubsection}%
  {3}%
  {}%
  {-0.2\baselineskip}%
  {0.1\baselineskip}%
  {}}%
\newcommand{\unnumberedsection}[1]{
    \@startsection{section}{1}{2.7cm}{-0.1\baselineskip}{0.1\baselineskip}{\normalfont\footnotesize\bfseries}*{#1}
}

\g@addto@macro\normalsize{%
  \setlength\abovedisplayskip{3pt plus 2pt minus 1pt}
  \setlength\belowdisplayskip{3pt plus 2pt minus 1pt}
  \setlength\abovedisplayshortskip{2pt plus 2pt minus 1pt}
  \setlength\belowdisplayshortskip{2pt plus 2pt minus 1pt}
}

\DeclareMathOperator*{\argmax}{arg\,max}
\title{Label-Context-Dependent Internal Language Model Estimation for CTC}

\author[affiliation={1}]{Zijian}{Yang$^*$}
\author[affiliation={1}]{Minh-Nghia}{Phan$^*$}
\author[affiliation={1,2}]{Ralf}{Schlüter}
\author[affiliation={1,2}]{Hermann}{Ney}
\affiliation{Human Language Technology and Pattern Recognition, Computer Science Department}{RWTH Aachen University}{Germany}
\affiliation{}{AppTek GmbH}{Germany}
\email{\{zyang, schlueter, ney\}@ml.rwth-aachen.de, minh.nghia.phan@rwth-aachen.de}

\usepackage{comment}

\begin{document}
\maketitle

\begin{abstract}
Although connectionist temporal classification (CTC) has the label context independence assumption, it can still implicitly learn a context-dependent internal language model (ILM) due to modern powerful encoders. In this work, we investigate the implicit context dependency modeled in the ILM of CTC. To this end, we propose novel context-dependent ILM estimation methods for CTC based on knowledge distillation (KD) with theoretical justifications. Furthermore, we introduce two regularization methods for KD. We conduct experiments on Librispeech and TED-LIUM Release 2 datasets for in-domain and cross-domain evaluation, respectively. Experimental results show that context-dependent ILMs outperform the context-independent priors in cross-domain evaluation, indicating that CTC learns a context-dependent ILM. The proposed label-level KD with smoothing method surpasses other ILM estimation approaches, with more than 13\% relative improvement in word error rate compared to shallow fusion. 
% The effect of context length and the combination of frame-level prior and context-dependent ILMs are also investigated.
% In more detail, we employ CTC as the teacher model and a standard LM as the student model, to distill CTC output probabilities on label and sequence levels. 
\end{abstract}

\section{Introduction \& Related Work}
\blfootnote{$*$ denotes equal contribution}
In automatic speech recognition, sequence-to-sequence (seq2seq) models have drawn more and more attention in recent years, due to their simplified training/decoding pipeline and strong performance. The most famous architectures include attention-based encoder-decoder models (AED) \cite{Tske2020SingleHA}, recurrent neural network transducers (RNN-T) \cite{graves2012sequence} and connectionist temporal classification (CTC) \cite{graves2006connectionist}. Although seq2seq models can be used as standalone models for decoding, integrating an external language model (ELM) trained on large-scale text data often leads to better performance for both in- and cross-domain tasks. Among various integration methods, Shallow Fusion (SF) is a widely used approach, combining the ELM and acoustic model (AM) probabilities in a log-linear manner.

Since a seq2seq model provides posterior probabilities for sequences, integrating it with an ELM during decoding requires compensation for the prior/internal language model (ILM) of the ASR model according to Bayes rule. While ILM estimation methods have been extensively studied for models with explicit label context dependency, such as AED and RNN-T \cite{meng2021internal, zeyer2021librispeech, zeineldeen2021investigating, zhou2022language}, due to the encoder-only architecture and the label context independence assumption, estimating the ILM of CTC is inherently challenging. Most existing ILM estimation approaches for CTC rely on heuristics. Das et al. proposed a heuristic ILM estimation method based on masking out the acoustic input and accumulating log-posteriors for only the masked time frames \cite{das2023mask}. Zhao et al. proposed to apply different masking rates for the acoustic input to detect the strength of the ILM of CTC \cite{zhaoregarding}. Other works apply the marginal frame-level prior as the ILM based on the context-independent assumption \cite{miao2015eesen, manohar2015semi}. However, due to powerful architectures applied in modern models, a label-context-dependent ILM can be learned in the encoder implicitly. In \cite{kanda2016maximum}, the authors showed that the bigram ILM outperformed the unigram ILM in search, indicating that the ILM of CTC is context-dependent. Nevertheless, \cite{kanda2016maximum} estimates the ILM using transcription text, which is decorrelated from the CTC output. 

In this work, we aim to address the implicit context dependence of the CTC ILM. Since CTC does not have a language-model-like module, we propose novel label-context-dependent ILM estimation methods for CTC based on label-level and sequence-level knowledge distillation (KD) with theoretical justifications. More precisely, we employ the CTC model as the teacher model, and a small ILM estimator as the student model, to distill CTC probabilities to the ILM estimator on label and sequence levels. To the best of our knowledge, this is the first context-dependent ILM estimation method based on CTC outputs. For the label-level distillation, we propose to compute the label posterior via CTC prefix probabilities. Moreover, we introduce two regularization methods for KD, i.e. smoothing and masking. Systematic comparisons are conducted for both in-domain and cross-domain evaluation on Librispeech (LBS) \cite{panayotov2015librispeech} and TEDLIUM Release 2 (Tedlium2) \cite{rousseau2014enhancing} datasets, respectively. Experimental results on Tedlium2 demonstrate that ILMs with context dependence work better than the context-independent priors in cross-domain evaluation, indicating a context-dependent ILM of CTC. Furthermore, in cross-domain evaluation, our proposed methods outperform previous ILM estimation methods with over 13\% relative improvement in word error rate compared to shallow fusion (SF). We also investigate the effect of combining frame-level prior with other ILM estimation approaches, as well as the impact of different context lengths on ILM compensation performance in decoding.

\section{CTC}

Given an input sequence $X=x_1,x_2,\dots, x_{T'}$ and a label vocabulary $\mathcal{V}$, the probability of the label sequence $a_1^S$ with $a_s \in \mathcal{V}$ is modeled by CTC as follows:\\
\scalebox{0.95}{\parbox{1.05\linewidth}{
\begin{align}
    P(a_1^S|X) &= \sum_{y_1^T: \mathcal{B}(y_1^T)=a_1^S} P(y_1^T |h_1^T) \notag \\
    & = \sum_{y_1^T: \mathcal{B}(y_1^T)=a_1^S} \prod_{t=1}^T P(y_t|h_t), \label{eq:ctcdef}
\end{align}
}}
where $y_1^T$ is the blank-augmented alignment sequence, $\mathcal{B}$ is the collapse function and $h_1^T$ is the sequence of encoder output vectors with potential subsampling. As shown in the decomposition in Eq. (\ref{eq:ctcdef}), the output of CTC has no explicit label context dependency.
% , which makes it possible to only have an encoder and a linear-softmax layer for CTC.
However, as a seq2seq model, CTC models the posterior probability of the label sequence. Therefore, when integrating an external language model (ELM), the correction for the internal language model (ILM) should be considered based on Bayes' theorem.\\
\scalebox{0.95}{\parbox{1.05\linewidth}{
\begin{equation}
    X \rightarrow {a^*}_1^{S^*} = \argmax_{S,a_1^S} P(a_1^S|X)\frac{ P^{\lambda_1}_{\text{ELM}}(a_1^S)}{P^{\lambda_2}_{\text{ILM}}(a_1^S)} 
    \label{eq:sf}
\end{equation}
}}
Here, $P_{\text{ELM}}(a_1^S)$ is the ELM probability, and $P_{\text{ILM}}(a_1^S)$ is the ILM probability of CTC. $\lambda_1$ and $\lambda_2$ are scales for the ELM and ILM, repectively. In practice, Viterbi approximation is usually applied for efficiency, which takes the best path during decoding. The ILM of CTC is defined as:
\begin{equation}
    P_{\text{ILM}}(a_1^S) := \sum_{X} Pr(X) P(a_1^S|X),
    \label{eq:ilmdef}
\end{equation}
where $Pr(X)$ denotes the true marginal distribution of the data. Since the true distribution is unknown, and computing the summation for all label sequences $a_1^S$ is intractable, in practice, $P_\text{ILM}$ is estimated by various approaches.

\section{Internal Language Model Estimation for CTC}

\subsection{Frame-Level Unigram Prior Estimation}
\label{sec:frameprior}
 Label prior is a simple estimation of the ILM. Manohar et al. proposed to compute the label prior by marginalizing the model posterior over all acoustic inputs \cite{manohar2015semi}. For CTC, this approach provides a context-independent frame-level prior (FP) estimation.\\
\scalebox{0.95}{\parbox{1.05\linewidth}{
\begin{align}
P_\text{FP}(y) = \frac{1}{\sum_{n=1}^N T_{n}}\sum_{n=1}^N \sum_{t=1}^{T_n}P(y|h_{n,t})
\end{align}
}}
Here, $N$ denotes the number of training sequences, $T_n$ denotes the encoder output length of sequence $n$, and $h_{n,t}$ denotes the hidden state of sequence $n$ at time frame $t$. Since this unigram is estimated on frame level, it is applied to the CTC output for each frame during decoding, rather than only applied when a new label is emitted. That is, we use the following $q(y_t,h_t)$ with a prior scale $\lambda_3$ to replace the original CTC output $p(y_t|h_t)$ when integrating with an during in decoding.\\
\scalebox{0.95}{\parbox{1.05\linewidth}{
\vspace{1em}
\begin{equation*}
    q(y_t,h_t) = \frac{P(y_t|h_t)}{P^{\lambda_3}_{\text{FP}}(y_t)}
\end{equation*}}}
\subsection{ILM Estimation with Knowledge Distillation}
 \label{sec:ILMKD}
We propose to utilize knowledge distillation (KD) to estimate a context-dependent ILM of CTC. More precisely, we employ CTC as the teacher model and a standard autoregressive LM $q_\theta(a_1^S)$ (ILM estimator) as the student model, allowing the LM to learn the ILM probability distribution from CTC. Since CTC does not explicitly model context dependencies, we assume that its ILM is relatively weak. Therefore, we employ a small LSTM language model as the ILM estimator in practice.

\subsubsection{Label-Level Knowledge Distillation for ILM estimation}
% To simplify notations, in the following discussion, we refer to $a_1^S$ as a complete label sequence, i.e. $a_1,\dots, a_S$ followed by \textless EOS\textgreater, and $a_1^s$ as a prefix $a_1,\dots, a_s$ without \textless EOS\textgreater\ following.
The label posterior of ILM can be computed as follows:\\
\scalebox{0.95}{\parbox{1.05\linewidth}{
\begin{align}
    P_\text{ILM}(a_s|a_1^{s-1}) &= \sum_{X} P(a_s,X|a_1^{s-1})\notag \\
    &\approx \sum_{X} P(a_s|a_1^{s-1},X) Pr(X|a_1^{s-1}) \notag \\
    & =\sum_{X} P(a_s|a_1^{s-1},X) \frac{Pr(X,a_1^{s-1})}{Pr(a_1^{s-1})} \label{eq:labelilmapprox}
\end{align}
}}
Although CTC does not model the label posterior probability $P(a_s|a_1^{s-1}, X)$ directly, it can be computed as follows:\\
\scalebox{0.95}{\parbox{1.05\linewidth}{
\begin{equation}
    P(a_s|a_1^{s-1}, X) = \frac{P(a_1^{s}, ...|X)}{P(a_1^{s-1},...|X)},
    \label{eq:ctcdecomp}
\end{equation}
}}
where $P(a_1^{s},...|X)$ is the CTC prefix probability defined as the marginalized probability of all label sequences that have $a_1^s$ as their prefix \cite{phdgraves,watanabe2017hybrid}.\\
\scalebox{0.95}{\parbox{1.05\linewidth}{
\begin{equation*}
    P(a_1^{s}, ...|X) = \sum_{v \in \big(\bigcup_{i=1}^{\infty} \mathcal{V}^i \cup \emptyset \big)} P(a_1^s \cdot v|X)
\end{equation*}
}}
While the end-of-sentence (\textless EOS\textgreater) token is crucial for an autoregressive LM to ensure proper normalization, CTC does not explicitly model the probability of \textless EOS\textgreater. However, CTC implicitly accounts for the \textless EOS\textgreater\ probability when computing the probability of the whole sequence. Therefore, the posterior probability of \textless EOS\textgreater\ can be computed via:\\
\scalebox{0.95}{\parbox{1.05\linewidth}{
\begin{equation*}
    P(\text{\textless EOS\textgreater}|a_1^{s}, X) = \frac{P(a_1^{s}|X)}{P(a_1^{s},...|X)}
\end{equation*}
}}
% For each training pair $(X, a_1^S)$, the training criterion is defined via the following Kullback–Leibler divergence:\\
% \scalebox{0.92}{\parbox{1.08\linewidth}{
% \begin{equation}
%     F(\theta_{\text{ILM}}, X, a_1^S) = \sum_{s=1}^{S+1} \sum_{a\in \mathcal{V}^+}P(a|a_1^{s-1}, X) \log \frac{P(a|a_1^{s-1}, X)}{P_\text{ILM}(a|a_1^{s-1}; \theta_\text{ILM})}
%     \label{eq:standardcriterion}
% \end{equation}
% }}
During training, We reuse the CTC training data to estimate the ILM. Let $(X_n, a_{n,1}^{S_n})$ be the $n$-th pair in the training data, and $N$ be the number of training pairs. The training criterion based on Kullback–Leibler divergence is defined as follows:\\
\scalebox{0.94}{\parbox{1.06\linewidth}{
% \begin{align}
%     F(\theta_\text{ILM}) &= \frac{1}{N}\sum_{n=1}^N F(\theta_{\text{ILM}}, X_n, a_{n,1}^{S_n}) \notag \\ 
%     & = \sum_{X,S, a_1^S} \Tilde{Pr}(X, a_1^S) F(\theta_{\text{ILM}}, X, a_1^S), \label{eq:labelstandard}
% \end{align}
\begin{align}
    &F(\theta) = \frac{1}{N}\sum_{n=1}^N  \sum_{s=1}^{S_n+1} \sum_{a\in \mathcal{V}^+}P(a|a_{n,1}^{s-1}, X_n) \log \frac{P(a|a_{n,1}^{s-1}, X_n)}{q_\theta(a|a_{n,1}^{s-1})} \label{eq:labelstandardindex} \\ 
    & = \sum_{X,S, a_1^S} \Tilde{Pr}(X, a_1^S)  \sum_{s=1}^{S+1} \sum_{a\in \mathcal{V}^+}P(a|a_1^{s-1}, X) \log \frac{P(a|a_1^{s-1}, X)}{q_\theta(a|a_1^{s-1})}, \label{eq:labelstandard}
\end{align}
}}
where $\mathcal{V}^+$ is the \textless EOS\textgreater-augmented vocabulary. The label posterior $P(a|a_{n,1}^{s-1}, X)$ is derived from the CTC output via Eq. \eqref{eq:ctcdecomp} and serves as the fixed teacher model. Meanwhile, $q_\theta(a|a_{n,1}^{s-1})$ is the output of a standard LM with learnable parameters $\theta$, applied as the ILM estimator for the ILM of CTC. Eq. \eqref{eq:labelstandard} is derived from Eq. \eqref{eq:labelstandardindex} by representing the training data with the empirical distribution $\Tilde{Pr}$. The motivation of this training criterion is: if there is sufficient amount of training data, i.e. $\Tilde{Pr}(X, a_1^S) \approx Pr(X, a_1^S)$, it can be proven that the label-level ILM approximation in Eq. (\ref{eq:labelilmapprox}) is exactly the global optimum $\hat{q}$ of the training criterion (\ref{eq:labelstandard}), i.e. $\hat{q}(a|a_1^{s-1}) \approx P_\text{ILM} (a|a_1^{s-1})$. A detailed proof is derived in Appendix A.

% \begin{align}
%     &F(\theta_\text{ILM}) = \sum_{X, a_1^S} Pr(X, a_1^S) F(\theta_{\text{ILM}}, X, a_1^S) \notag \\
%     & = \sum_{X, a_1^S} Pr(X, a_1^S) \sum_{s=1}^S \sum_{a\in \mathcal{V}}P(a|a_1^{s-1}, X) \log \frac{P(a|a_1^{s-1}, X)}{P_\text{ILM}(a|a_1^{s-1}; \theta_\text{ILM})} \notag \\
%     & = \sum_{X, s, a_1^{s-1}}Pr(X, a_1^{s-1}) \sum_{a\in \mathcal{V}}P(a|a_1^{s-1}, X) \log \frac{P(a|a_1^{s-1}, X)}{P_\text{ILM}(a|a_1^{s-1}; \theta_\text{ILM})} \notag \\
%     &= \sum_{s,a_1^{s-1}} Pr(a_1^{s-1}) \sum_{a \in \mathcal{V}}\sum_{X} P(a|a_1^{s-1}, X) Pr(X|a_1^{s-1}) \log \frac{P(a|a_1^{s-1}, X)}{P_\text{ILM}(a|a_1^{s-1}; \theta_\text{ILM})}
% \end{align}   
In practice, the available training data is limited, and achieving a global optimum is not guaranteed. Since CTC training data is used for ILM estimation, the CTC model may become overconfident on previously seen data. We observe that the posterior probability is usually close to 1 for the ground truth label in training data and 0 for other labels, which makes this criterion close to the standard cross-entropy criterion for LM training on transcriptions. To mitigate this issue, we propose two regularization methods for ILM estimator training: KD with smoothing and masking.\\
\textbf{Label-level KD with smoothing:}
\label{sec:labelsampling}
Due to the limited amount of training data, the empirical distribution $\Tilde{Pr}$ is typically sparse. Inspired by some count-based LM smoothing methods like absolute discounting and Jelinek–Mercer smoothing, we propose to smooth the empirical distribution $\Tilde{Pr}(X,a_1^S)$ in Eq. \eqref{eq:labelstandard} by interpolating the marginal distributions $\Tilde{Pr}(X)$ and $\Tilde{Pr}(a_1^S)$:\\
\scalebox{0.95}{\parbox{1.05\linewidth}{
\begin{equation*}
    \overline{Pr}(X,a_1^S) = \alpha \Tilde{Pr}(X,a_1^S) + (1-\alpha)\Tilde{Pr}(X)\Tilde{Pr}(a_1^S),
\end{equation*}
}}
where $\alpha$ is an interpolation factor. The latter smoothing term can be regarded as dropping the interdependencies between $X$ and $a_1^S$. The training criterion is then defined as:\\
\scalebox{0.91}{\parbox{1.09\linewidth}{
% \begin{align}
%     F_\text{smoothing}(\theta_\text{ILM}, \mathfrak{B}) = \sum_{b=1}^{|\mathfrak{B}|} \sum_{b'=1}^{|\mathfrak{B}|} \beta(b,b') F(\theta_\text{ILM}, X_{b'}, a_{b,1}^{S_b})\\
%     \label{eq:labelsampling}
% \end{align}
\begin{align}
    &F_\text{smoothing}(\theta) \notag \\
     &= \sum_{X,S, a_1^S} \overline{Pr}(X, a_1^S)  \sum_{s=1}^{S+1} \sum_{a\in \mathcal{V}^+}P(a|a_1^{s-1}, X) \log \frac{P(a|a_1^{s-1}, X)}{q_\theta(a|a_1^{s-1})} \label{eq:labelsamplinginter}\\
    &=\sum_{n=1}^{N} \sum_{n'=1}^{N} \frac{\beta(n,n')}{N}\sum_{s=1}^{S_n+1} \sum_{a\in \mathcal{V}^+}P(a|a_{n,1}^{s-1}, X_{n'}) \log \frac{P(a|a_{n,1}^{s-1}, X_{n'})}{q_\theta(a|a_{n,1}^{s-1})} 
\label{eq:labelsampling}
\end{align}
}}
\\
% Here, $(X_n, a_{n,1}^{S_n}) \in \mathfrak{B}$ is the $n$-th training pair in the batch, with the label sequence length $S_n$, and 
Here, $\beta(n,n')$ is defined by Kronecker delta $\delta$:\\
\scalebox{0.95}{\parbox{1.05\linewidth}{
\begin{equation*}
    % \beta(n,n') = \left \{\begin{array}{ll}
    % \alpha + \frac{1-\alpha}{N},     &  \text{if } n=n'\\
    % \frac{1-\alpha}{N},     & \text{otherwise}
    % \end{array}
    % \right .
    \beta(n,n') = \delta(n,n') \alpha + \frac{1-\alpha}{N},
\end{equation*}
}}
The equivalence between Eq. \eqref{eq:labelsamplinginter} and Eq. \eqref{eq:labelsampling} is derived in Appendix B. In general, for a label sequence $a_{n,1}^{S_n}$ in the training data, the smoothing criterion computes an averaged probability over all inputs, rather than only considering the pairwise input $X_n$ like in Eq. \eqref{eq:labelstandardindex}.
In practice, since the smoothing over the whole training data is infeasible, we conduct the smoothing within a mini-batch.
\\
\textbf{Label-level KD with Masking}
To mitigate the overconfidence of CTC, we propose to mask the acoustic evidence of the label based on alignments. Let $t_1^S$ be (sub-)word boundaries provided by a GMM alignment, where $t_{s}$ is the end time frame of label $s$. Each label position $s$ has a probability $p_\text{mask}$ to mask out the corresponding acoustic input $x_{t_{s-1}+1}^{t_{s}}$. For training efficiency, we mask out multiple positions for each training pair. Let the set of masked positions for the $n$-th sequence be $\mathcal{M}_n \subseteq  \{1, 2, \dots, S_n\}$,  the training criterion is defined as:\\
\scalebox{0.95}{\parbox{1.05\linewidth}{
\begin{align*}
    &F_\text{masking}(\theta)\\
    &= \frac{1}{N}\sum_{n=1}^N \sum_{s\in \mathcal{M}_n} \sum_{a \in \mathcal{V}^+} P(a|a_{n,1}^{s-1}, \overline{X}_n)\log \frac{P(a|a_{n,1}^{s-1}, \overline{X}_n)}{q_\theta(a|a_{n,1}^{s-1})}
\end{align*}
}}
where $\overline{X}_n$ is the masked input sequence.
\subsubsection{Sequence-Level Knowledge Distillation for ILM estimation}
The KD from CTC to ILM estimator can also be done on the sequence level. Following the idea of the smoothing method in Sec. \ref{sec:labelsampling}, we smooth the empirical distribution for sequence-level KD:\\
\scalebox{0.94}{\parbox{1.06\linewidth}{
\begin{align}
    F_{\text{seq}}(\theta) =  \sum_{n=1}^{N} \sum_{n'=1}^{N} \frac{\beta(n,n')}{N}P(a_{n,1}^{S_n}|X_{n'}) \log \frac{P(a_{n,1}^{S_n}|X_{n'})}{q_\theta(a_{n,1}^{S_n})}
    \label{eq:seqcriterion}
\end{align}
}}
The main difference between the sequence-level KD criterion \eqref{eq:seqcriterion} and the label-level KD criterion \eqref{eq:labelsampling} is that \eqref{eq:seqcriterion} transfers sequence probabilities to the ILM estimator, while \eqref{eq:labelsampling} transfers label posterior probabilities.

\section{Experiments}
\subsection{Setup}
Experiments are done on the 960 hours Librispeech (LBS) dataset \cite{panayotov2015librispeech} for in-domain evaluation and TED-LIUM Release 2 (Tedlium2) dataset \cite{rousseau2014enhancing} for cross-domain evaluation using the RETURNN framework \cite{zeyer2018returnn} based on Pytorch \cite{paszke2019pytorch}. We use byte-pair-encoding (BPE) \cite{sennrich2015neural} as output labels with a vocabulary size of 10k.

We use a 12-layer conformer \cite{gulati20_interspeech} CTC model as our baseline acoustic model. The hidden dimension of the conformer block is 512, and the feed-forward dimension is 2048. We employ 80-dimensional log Mel features as the input to CTC. SpecAugment is applied for data augmentation in CTC training. A downsampling factor of 6 is applied. The model is trained for 100 full epochs on LBS with AdamW optimizer \cite{loshchilov2017decoupled} and one-cycle learning rate schedule \cite{saon2021advancing,xu24_interspeech}, with a peak learning rate of 1e-3. This CTC model is also applied as the teacher model in KD for ILM estimation.
We assume that the ILM of CTC is relatively weak. Therefore, we employ a one-layer LSTM LM as the ILM estimator. The embedding dimension is 128, and the LSTM hidden dimension is 1000. ILM estimators are trained via criteria proposed in Sec. \ref{sec:ILMKD}. When computing teacher probabilities, SpecAugment is disabled to reflect the exact output of CTC. We train ILM estimators on the LBS transcriptions for 5 full epochs. We also train a standard LM with the same architecture as the ILM estimator on LBS transcriptions for 30 full epochs, referred to as the transcription LM.

During decoding, we employ a 24-layer transformer LM \cite{Irie2019LanguageMW} as the ELM for in-domain evaluation and a 4-layer LSTM LM as the ELM for cross-domain evaluation, with perplexity (PPL) 37 and 48 on dev sets, respectively. We apply the time-synchronous Viterbi search implemented in RETURNN. All the scales applied in search are optimized on the dev sets. Since we observed that the ILM PPL is not correlated with the word error rate (WER), we select the ILM estimator checkpoints based on evaluation results on dev sets. The smoothing factor $\alpha$ and the masking rate $p_\text{mask}$ are also tuned based on recognition results on dev sets. We use $\alpha = 0.5$ for smoothing and $p_{\text{mask}}=0.4$ for masking in our experiments. Our code can be found online.\footnote{https://github.com/rwth-i6/returnn-experiments/tree/master/2025-label-dependent-ILM-ctc}
\subsection{In-Domain and Cross-Domain Evaluation}
Table \ref{tab:mainresult} shows evaluation results for the in-domain Librispeech and cross-domain Tedlium2 Corpora. The unigram ILM is computed via renormalizing FP without the blank label and applied on the label level. Similar to \cite{kanda2016maximum}, the transcription LM served as an approximation of the ILM of CTC. As expected, integrating the ELM brings significant improvements over the standalone CTC, and all ILM correction approaches further improve the performance compared to SF. Moreover, different ILM estimation methods perform comparably on in-domain tasks, but differ more on cross-domain tasks. We assume that this is because ILM correction has two functions: boosting label emission and rebalancing the label distribution, consistent with discussions on ILM in \cite{zhou2022language,yang2024relation}. For in-domain tasks, since the ILM of CTC is relatively weak, ILM correction mainly serves to boost the label emission, while precise ILM estimation for the label distribution reshaping is unnecessary. Therefore, all ILM estimation methods perform similarly. However, for cross-domain tasks, because of the mismatch between the underlying distribution of the source and target domain, proper ILM estimation for label distribution rebalancing is important. 

In cross-domain evaluation results, we observe that all the context-dependent ILM estimation methods outperform the context-independent priors (unigram and FP), indicating that CTC learns a context-dependent ILM implicitly due to the powerful encoder. The label-level KD without any regularization performs similar to the transcription LM, which verifies the statement in Sec. \ref{sec:ILMKD} that without regularization, the estimated ILM would be close to a transcription LM. Moreover, our proposed label-level KD methods with regularization (smoothing and masking) surpass other methods, indicating better ILM estimation and the necessity of regularization. Meanwhile, the sequence-level KD is worse than the label-level KD. This may be because label-level KD provides more information about the CTC output distribution, as it considers the probabilities of all labels in the vocabulary for a given prefix in the training data, whereas sequence-level distillation only accounts for the probabilities of sequences present in the training data. Overall, our proposed label-level KD with smoothing achieves the best results in cross-domain evaluation, yielding a relative improvement of over 13\% compared to SF. We also compute PPLs of the estimated ILMs. Unlike the known correlation between the PPL of the ELM and WER \cite{klakow2002testing, irie2020advancing}, we do not observe a correlation between ILM PPLs and WERs. This suggests that ILM estimator checkpoints should not be selected based on the PPL, and new metrics need to be explored to better evaluate ILMs.

\begin{table}[t!]
\caption{Recognition results for the in-domain Librispeech and cross-domain Tedlium2 Corpora. The ILM regularization is done during the ILM estimator training. The ILM PPLs are computed on the dev sets. The unigram ILM is computed via renormalizing frame-level prior (FP) without the blank label. Trans LM refers to applying transcription LM as an approximation of the ILM. Label and seq KD refer to label- and sequence-level KD methods for ILM estimation proposed in Sec. \ref{sec:ILMKD}.}
\label{tab:mainresult}
\scalebox{0.82}{\parbox{1.18\linewidth}{
\setlength{\tabcolsep}{0.2em}

\begin{tabular}{|lcc|ccccc|ccc|}
\hline
\multicolumn{1}{|c|}{\multirow{4}{*}{ELM}}&  \multicolumn{1}{c|}{\multirow{4}{*}{ILM}} & \multicolumn{1}{c|}{\multirow{4}{*}{\begin{tabular}[c]{@{}c@{}}ILM reg-\\ ularization\end{tabular}}}    & \multicolumn{5}{c|}{Librispeech}                                                                                                                                                       & \multicolumn{3}{c|}{Tedlium2}                                                                                                                               \\ \cline{4-11} 
 \multicolumn{1}{|c|}{}         &       \multicolumn{1}{c|}{}          &   & \multicolumn{1}{c|}{\multirow{3}{*}{\begin{tabular}[c]{@{}c@{}}ILM\\ PPL\end{tabular}}} & \multicolumn{4}{c|}{WER{[}\%{]}}                                                                     & \multicolumn{1}{c|}{\multirow{3}{*}{\begin{tabular}[c]{@{}c@{}}ILM\\ PPL\end{tabular}}} & \multicolumn{2}{c|}{WER{[}\%{]}}                                          \\ \cline{5-8} \cline{10-11} 
 \multicolumn{1}{|c|}{}                        & \multicolumn{1}{c|}{}   & \multicolumn{1}{c|}{}                   &    \multicolumn{1}{c|}{}                                             & \multicolumn{2}{c|}{dev}                                & \multicolumn{2}{c|}{test}          & \multicolumn{1}{c|}{}                                                                   & \multicolumn{1}{c|}{\multirow{2}{*}{dev}} & \multirow{2}{*}{test} \\ \cline{5-8}
 \multicolumn{1}{|c|}{}                        & \multicolumn{1}{c|}{}& \multicolumn{1}{c|}{}    & \multicolumn{1}{c|}{}                                                                   & \multicolumn{1}{c|}{clean} & \multicolumn{1}{c|}{other} & \multicolumn{1}{c|}{clean} & other & \multicolumn{1}{c|}{}                                                                   & \multicolumn{1}{c|}{}                     &                       \\ \hline
\multicolumn{1}{|c|}{no} &\multicolumn{1}{c|}{\multirow{2}{*}{no}}   & \multicolumn{1}{c|}{\multirow{6}{*}{-}}                     & \multicolumn{1}{c|}{-}                                                                   & \multicolumn{1}{c|}{3.0}      & \multicolumn{1}{c|}{6.8}      & \multicolumn{1}{c|}{3.2}      &   7.2    & \multicolumn{1}{c|}{-}                                                                   & \multicolumn{1}{c|}{17.7}                     &         18.7              \\ \cline{1-1} \cline{4-11}
\multicolumn{1}{|c|}{\multirow{8}{*}{yes}}& \multicolumn{1}{l|}{}     &                      & \multicolumn{1}{c|}{-}                                                                   & \multicolumn{1}{c|}{2.2}      & \multicolumn{1}{c|}{4.8}      & \multicolumn{1}{c|}{2.2}      &    5.3   & \multicolumn{1}{c|}{-}                                                                   & \multicolumn{1}{c|}{14.3}                     &         15.9              \\ \cline{2-2} \cline{4-11}
\multicolumn{1}{|c|}{}& \multicolumn{1}{c|}{FP}     &    & \multicolumn{1}{c|}{-}                                                                   & \multicolumn{1}{c|}{2.1}      & \multicolumn{1}{c|}{4.4}      & \multicolumn{1}{c|}{2.1}      &    4.9   & \multicolumn{1}{c|}{-}                                                                   & \multicolumn{1}{c|}{13.2}                     &           14.7            \\ \cline{2-2} \cline{4-11}
\multicolumn{1}{|c|}{}& \multicolumn{1}{c|}{unigram}     &      & \multicolumn{1}{c|}{ 1223}                                                                    & \multicolumn{1}{c|}{2.1 }      & \multicolumn{1}{c|}{ 4.4}      & \multicolumn{1}{c|}{ 2.1}      &   4.9   & \multicolumn{1}{c|}{1307}                                                                   & \multicolumn{1}{c|}{12.9 }                     &    14.9                   \\ \cline{2-2} \cline{4-11}
\multicolumn{1}{|c|}{}& \multicolumn{1}{c|}{trans LM}     &      & \multicolumn{1}{c|}{145}                                                                    & \multicolumn{1}{c|}{2.1}      & \multicolumn{1}{c|}{4.4}      & \multicolumn{1}{c|}{2.1}      &  4.9     & \multicolumn{1}{c|}{278}                                                                   & \multicolumn{1}{c|}{12.4}                     &       14.5                \\ \cline{2-2} \cline{4-11}
\multicolumn{1}{|c|}{} & \multicolumn{1}{l|}{\multirow{3}{*}{label KD}} &           & \multicolumn{1}{c|}{140}                                                                   & \multicolumn{1}{c|}{2.1}      & \multicolumn{1}{c|}{4.4}      & \multicolumn{1}{c|}{2.1}      &   4.9    & \multicolumn{1}{c|}{255}                                                                   & \multicolumn{1}{c|}{12.4}                     &             14.4          \\ \cline{3-11}
\multicolumn{1}{|l|}{}& \multicolumn{1}{l|}{}  & masking & \multicolumn{1}{c|}{169}                                                                   & \multicolumn{1}{c|}{2.0}      & \multicolumn{1}{c|}{\textbf{4.3}}      & \multicolumn{1}{c|}{2.0}      &   4.8    & \multicolumn{1}{c|}{286}                                                                   & \multicolumn{1}{c|}{12.2}                     &       14.0                \\ \cline{3-11}
\multicolumn{1}{|c|}{} & \multicolumn{1}{l|}{}  & \multirow{2}{*}{smoothing} & \multicolumn{1}{c|}{191}                                                                   & \multicolumn{1}{c|}{\textbf{2.0}}      & \multicolumn{1}{c|}{4.4}      & \multicolumn{1}{c|}{\textbf{2.0}}      &    \textbf{4.8}   & \multicolumn{1}{c|}{297}                                                                   & \multicolumn{1}{c|}{\textbf{11.9}}                     &         \textbf{13.8}              \\ \cline{2-2} \cline{4-11}

\multicolumn{1}{|l|}{}& \multicolumn{1}{c|}{seq KD} &        & \multicolumn{1}{c|}{293}                                                                   & \multicolumn{1}{c|}{2.1}      & \multicolumn{1}{c|}{4.5}      & \multicolumn{1}{c|}{2.1}      &    5.0   & \multicolumn{1}{c|}{395}                                                                   & \multicolumn{1}{c|}{12.5}                     &              14.3         \\ \hline
\end{tabular}
}}
\end{table}
\begin{table}[t!]
\centering
\caption{WERs {[}\%{]} on Tedlium2 dev and test dataset for different ILM estimation methods combined with frame-level prior. The ELM is used in decoding. The ILM regularization is done during the ILM estimator training. Trans LM refers to applying transcription LM as an approximation of the ILM. Label and seq KD refer to label- and sequence-level KD methods for ILM estimation proposed in Sec. \ref{sec:ILMKD}.}
\label{tab:jointprior}

\scalebox{1.0}{\parbox{1.0\linewidth}{
\setlength{\tabcolsep}{0.2em}
\centering
\begin{tabular}{|lc|cc|cc|}
        \hline
    \multicolumn{1}{|c|}{\multirow{2}{*}{ILM}} & \multicolumn{1}{c|}{\multirow{2}{*}{\begin{tabular}[c]{@{}c@{}}ILM reg-\\ ularization\end{tabular}}} & \multicolumn{2}{c|}{dev}              & \multicolumn{2}{c|}{test}             \\ \cline{3-6} 
       \multicolumn{1}{|c|}{} & \multicolumn{1}{c|}{}                        & \multicolumn{1}{c|}{w/ FP} & w/o FP & \multicolumn{1}{c|}{w/ FP} & w/o FP \\ \hline
        \multicolumn{1}{|c|}{no} &\multicolumn{1}{c|}{-}                & \multicolumn{1}{c|}{13.2}  & 14.3     & \multicolumn{1}{c|}{14.7}  & 15.9     \\ \hline 
        \multicolumn{1}{|c|}{trans LM} &   \multicolumn{1}{c|}{\multirow{2}{*}{-}}        & \multicolumn{1}{c|}{12.0}  &  12.4    & \multicolumn{1}{c|}{13.9}  &   14.5   \\ \cline{1-1} \cline{3-6} 
        \multicolumn{1}{|c|}{\multirow{3}{*}{label KD}}  & \multicolumn{1}{c|}{}           & \multicolumn{1}{c|}{12.1}  & 12.4     & \multicolumn{1}{c|}{13.8}  & 14.4     \\ \cline{2-6}
                    \multicolumn{1}{|c|}{}                                       & \multicolumn{1}{c|}{masking}            & \multicolumn{1}{c|}{12.1}  &   12.2   & \multicolumn{1}{c|}{13.9}  &    14.0  \\ \cline{2-6}
        \multicolumn{1}{|c|}{}                             & \multicolumn{1}{c|}{\multirow{2}{*}{smoothing}}          & \multicolumn{1}{c|}{11.8} & 11.9 & \multicolumn{1}{c|}{13.7} & 13.8 \\ \cline{1-1} \cline{3-6} 

        \multicolumn{1}{|c|}{seq KD} &\multicolumn{1}{c|}{} & \multicolumn{1}{c|}{11.9} & 12.5 & \multicolumn{1}{c|}{14.1} & 14.3 \\ \hline 
    \end{tabular}
% }
}}
\end{table}

\subsection{Ablation Study}

\subsubsection{Using Frame-Level Prior and ILM Jointly}
As discussed in Sec. \ref{sec:frameprior}, the frame-level prior is applied at each time frame. Strictly speaking, this is not a standard ILM correction method, as ILM correction should be defined on label sequences. Therefore, in this section, we examine the effect of combining the ILM defined on labels with frame-level prior correction. Table \ref{tab:jointprior} shows the comparison between various ILM estimation methods with and without frame-level prior. It is observed the transcription LM achieves notable improvement when combined with the frame-level prior. This is because the transcription LM models only the distribution of text data and is decorrelated from the CTC output. Consequently, it benefits from the frame-level prior estimated from CTC outputs.
%We observe similar effects on the standard label-ILM estimation method without regularization.
Sequence-level KD also benefits from the local distribution of the frame-level prior, as it learns the probabilities of entire sequences from CTC, lacking local information. In contrast, label-level KD with smoothing and masking gains little from the frame-level prior. This indicates that ILM correction has a similar effect to frame-level prior correction, but our method provides a more accurate estimation of the ILM.

\subsubsection{ILM with Different Context Lengths}

We also investigate the effect of different context lengths for the ILM. We employ two-layer feedforward networks for limited-context ILM estimators. For transcription LMs, since the distribution is independent of the CTC output, no clear correlation is observed between context length and performance. For label-level ILM, we observe that ILM with context-6 performs slightly better than context-1 and context-10 but worse than full-context. This may be due to the feedforward network's limited ability to model long contexts (e.g., context-10), making context-10 less effective than context-6. Overall, the optimal context length for CTC ILM requires further investigation.
% another possibility, the training criterion is derived for full context.

\begin{table}[h!]
\centering
\caption{WERs {[}\%{]} on Tedlium2 dev and test dataset for transcription LM and label-level ILM estimation with sampling across different context lengths. ELM is applied in recognition, while frame-level prior is not applied. Transcription LM is applied as approximated ILM.}
\label{tab:contextlength}
\scalebox{0.95}{\parbox{1.05\linewidth}{
\setlength{\tabcolsep}{0.2em}
\centering
\begin{tabular}{|l|c|cc|}
        \hline
        \multicolumn{1}{|c|}{\multirow{2}{*}{ILM}} & \multirow{2}{*}{Context} & \multicolumn{2}{c|}{WER} \\ 
        \cline{3-4} 
        & & \multicolumn{1}{c|}{dev} & test \\ 
        \hline
        \multicolumn{1}{|c|}{-} & - & \multicolumn{1}{c|}{14.3}& 15.9 \\ \hline
        \multirow{4}{*}{transcription LM} 
        & 1  & \multicolumn{1}{c|}{12.4} & 14.3 \\ \cline{2-4} 
        & 6  & \multicolumn{1}{c|}{12.7} & 14.6 \\ \cline{2-4} 
        & 10 & \multicolumn{1}{c|}{12.8} & 14.4 \\ \cline{2-4} 
        & full context & \multicolumn{1}{c|}{12.4} & 14.5 \\
        \hline
        \multirow{4}{*}{label-level KD w/ smoothing} 
        & 1  & \multicolumn{1}{c|}{12.4} & 14.2 \\ \cline{2-4} 
        & 6  & \multicolumn{1}{c|}{12.2} & 14.2 \\ \cline{2-4} 
        & 10 & \multicolumn{1}{c|}{12.2} & 14.4 \\ \cline{2-4} 
        & full context & \multicolumn{1}{c|}{11.9} & 13.8 \\
        \hline
    \end{tabular}
}}
\end{table}
\vspace{-0.2cm}
\section{Conclusion}
In this work, we investigated the implicit context dependence of the internal language model (ILM) of connectionist temporal classification (CTC). To this end, we proposed novel ILM estimation methods for CTC based on knowledge distillation (KD). We provided a solid theoretical background for label- and sequence-level KD in the probability space. Moreover, two regularization methods were introduced to improve the performance of the estimated ILM. We conducted our experiments on Librispeech for in-domain evaluation and TEDLIUM Release 2 for cross-domain evaluation. Experimental results showed that in cross-domain tasks, context-dependent ILMs outperformed the context-independent priors, indicating that CTC inherently learned a context-dependent ILM. Moreover, our proposed label-level KD with sampling method achieved the best performance, demonstrating a relative improvement of over 13\% in word error rate compared to shallow fusion. Additionally, we investigated the impact of context length and the effect of combining frame-level prior and ILMs for label sequences, providing further insights into ILM estimation.
{
\begin{spacing}{0.5}
{\footnotesize
\unnumberedsection{Acknowledgments}
%\section{acknowledgments}
% \scriptsize

% \singlespacing}

\scriptsize
\selectfont
This work was partially supported by the project RESCALE within the
program \textit{AI Lighthouse Projects for the Environment, Climate,
Nature and Resources} funded by the Federal Ministry for the Environment,
Nature Conservation, Nuclear Safety and Consumer Protection (BMUV), funding
ID: 67KI32006A.
}
\end{spacing}
}

\bibliographystyle{IEEEtran}
\bibliography{mybib}
\newpage
\appendix
{
\Huge{Appendix}
}
\section{The Global Optimum of Criterion \eqref{eq:labelstandard}}
In this section, we derive the global optimum of the training criterion \eqref{eq:labelstandard}, and show that this is equivalent to the ILM definition in Eq. \eqref{eq:labelilmapprox}. In order to distinguish a complete sequence and a prefix, we write out the \textless EOS \textgreater \quad label explicitly. To simplify the discussion, we assume that the maximum sequence length is finite. Therefore, all summations are done over a finite set, and exchanging the order of summations is always possible. Assume an unlimited amount of training data so that empirical distributions are the same as true distributions, i.e. $\Tilde{Pr}(\cdot) = Pr(\cdot)$. We have:\\
\scalebox{0.9}{\parbox{1.1\linewidth}{
\begin{align}
    &F(\theta) =  \notag \\
    & = \sum_{S, X, a_1^S } Pr(X, a_1^S\cdot \text{\textless EOS \textgreater}) \sum_{s=1}^{S+1} \sum_{a\in \mathcal{V}^{+}}P(a|a_1^{s-1}, X) \log \frac{P(a|a_1^{s-1}, X)}{q_{\theta}(a|a_1^{s-1})} \notag \\
    & = \sum_{S}\sum_{s=1}^{S+1}\sum_{X, a_1^S}Pr(X, a_1^S\cdot \text{\textless EOS \textgreater})  \sum_{a\in \mathcal{V}^{+}}P(a|a_1^{s-1}, X) \log \frac{P(a|a_1^{s-1}, X)}{q_{\theta}(a|a_1^{s-1})} \notag \\
    & = \sum_{S}\sum_{s=1}^{S+1}\sum_{X, a_1^{s-1}} Pr(X, a_1^{s-1}) \cdot \sum_{a_s^S} Pr(a_{s}^S \cdot \text{\textless EOS \textgreater}|a_1^{s-1},X)\notag  \\
    & \qquad \cdot \sum_{a\in \mathcal{V}^{+}}P(a|a_1^{s-1}, X) \log \frac{P(a|a_1^{s-1}, X)}{q_{\theta}(a|a_1^{s-1})} \notag \\
    & = \sum_{X,s,a_1^{s-1}} Pr(X, a_1^{s-1})\sum_{a\in \mathcal{V}^{+}}P(a|a_1^{s-1}, X) \log \frac{P(a|a_1^{s-1}, X)}{q_{\theta}(a|a_1^{s-1})} \notag \\
    & \quad \cdot \underbrace{\sum_{S\geq s-1, a_{s}^S}  Pr(a_{s}^S \cdot \text{\textless EOS \textgreater}|a_1^{s-1},X)}_{\begin{array}{c}=1 \text{ c.f. the normalization of } Pr(\cdot |a_1^{s-1},X),\\
    \quad a_{s}^{s-1} \text{refers to an empty sequence}\end{array} } \notag \\
    & = \sum_{X, s, a_1^{s-1}}Pr(X, a_1^{s-1}) \sum_{a\in \mathcal{V}^+}P(a|a_1^{s-1}, X) \log \frac{P(a|a_1^{s-1}, X)}{q_\theta(a|a_1^{s-1})} \notag \\
    &= \sum_{s,a_1^{s-1}} Pr(a_1^{s-1}) \sum_{a \in \mathcal{V}^+}\sum_{X} P(a|a_1^{s-1}, X) Pr(X|a_1^{s-1}) \log \frac{P(a|a_1^{s-1}, X)}{q_\theta(a|a_1^{s-1})}
    \label{eq:derivefinal}
\end{align}
}}
Since $\sum_{X} P(a|a_1^{s-1}, X) Pr(X|a_1^{s-1})$ is a properly normalized distribution over $a \in \mathcal{V}^+$, according to Gibbs' inequality, the global optimum of Eq. \eqref{eq:derivefinal} is obtained via:
\begin{equation*}
    \hat{q}_\theta(a|a_1^{s-1}) = \sum_{X} P(a|a_1^{s-1}, X) Pr(X|a_1^{s-1}), \forall a\in \mathcal{V}^+
\end{equation*}
Namely, equivalent to the definition Eq. \eqref{eq:labelilmapprox}

\section{Equivalence between Eq. \eqref{eq:labelsamplinginter} and \eqref{eq:labelsampling}}

Here, we derive that Eq. \eqref{eq:labelsamplinginter} is equivalent to Eq. \eqref{eq:labelsampling}. By definition, the empirical distributions are computed via:
\begin{align*}
    &\Tilde{Pr}(X,a_1^S) = \frac{1}{N} \sum_{n=1}^N \delta(X,X_n) \delta(a_1^S, a_{n,1}^{S_n}),\\
    &\Tilde{Pr}(X) =  \frac{1}{N} \sum_{n=1}^N \delta(X,X_n),\quad \Tilde{Pr}(a_1^S) = \frac{1}{N} \sum_{n=1}^N \delta(a_1^S, a_{n,1}^{S_n})
\end{align*}
To simplify notations, we define\\
\scalebox{0.85}{\parbox{1.15\linewidth}{
\begin{equation*}
    G(P, q|X,a_1^S):= \sum_{s=1}^{S+1}\sum_{a\in \mathcal{V}^+}P(a|a_1^{s-1}, X) \log \frac{P(a|a_1^{s-1}, X)}{q_\theta(a|a_1^{s-1})} 
\end{equation*}
}}
Therefore, Eq. \eqref{eq:labelsamplinginter} can be rewritten as:\\
\scalebox{0.85}{\parbox{1.15\linewidth}{
\begin{align*}
  &F(\theta) =   \sum_{X,S, a_1^S} \overline{Pr}(X, a_1^S)   G(P, q|X,a_1^S)\\
  & = \sum_{S,X,a_1^S}\frac{\alpha}{N} \sum_{n=1}^N \delta(a_1^S, a_{n,1}^{S_n})\delta(X,X_n)  G(P, q|X,a_1^S)\\
  & + \sum_{S,X,a_1^S}\frac{1-\alpha}{N} \sum_{n=1}^N \delta(a_1^S, a_{n,1}^{S_n})\frac{1}{N} \sum_{n'=1}^N \delta(X,X_{n'}) G(P, q|X,a_1^S)\\
  & = \frac{\alpha}{N} \sum_{n=1}^N \sum_{(X,a_1^S):X=X_n, a_1^S = a_{1,n}^{S_n}} G(P, q|X,a_1^S) \\
  & \quad + \frac{1-\alpha}{N}\sum_{n=1}^N \sum_{a_1^S: a_1^S = a_{1,n}^{S_n}} \frac{1}{N}\sum_{n'=1}^N \sum_{X: X = X_{n'}}G(P, q|X,a_1^S)\\
  & = \frac{\alpha}{N} \sum_{n=1}^N   G(P, q|X_n,a_{n,1}^{S_n}) + \frac{1}{N} \sum_{n=1}^N \sum_{n'=1}^N \frac{1-\alpha}{N}G(P, q|X_{n'},a_{n,1}^{S,n})\\
  & = \frac{1}{N}\sum_{n=1}^N \sum_{n'=1}^N \beta(n,n') G(P, q|X_{n'},a_{n,1}^{S_n})\\
  & =\sum_{n=1}^{N} \sum_{n'=1}^{N} \frac{\beta(n,n')}{N}\sum_{s=1}^{S_n+1} \sum_{a\in \mathcal{V}^+}P(a|a_{n,1}^{s-1}, X_{n'}) \log \frac{P(a|a_{n,1}^{s-1}, X_{n'})}{q_\theta(a|a_{n,1}^{s-1})} 
\end{align*}
}}
The derivation is done and we have shown that Eq. \eqref{eq:labelsamplinginter} and \eqref{eq:labelsampling} are equivalent.
\end{document}